# Gain-Embedded Meta Mirrors for Optically Pumped Semiconductor Disk Lasers

Zhou Yang, David Lidsky, and Mansoor Sheik-Bahae*
*Department of Physics and Astronomy*
University of New Mexico, Albuquerque, NM
*msb@unm.edu

**ABSTRACT:** We propose and analyze an active mirror structure that uses a subwavelength grating reflector combined with optical gain. The structure is designed to be directly bonded to a thermal substrate (such as diamond) for efficient heat removal. We present optical wave propagation and thermal transport analysis and show that such a structure is well suited for power scaling of optically pumped semiconductor disk lasers to multi-kilowatt CW power operation.

In recent years, optically pumped semiconductor (disk) lasers (OPSLs), also known as vertical-external-cavity surface-emitting lasers (VECSELs) [1], have emerged as highly versatile and compact solid-state laser sources with good beam quality for myriads of scientific and industrial applications [2][3][4]. Like most lasers, heat accumulation and the subsequent thermal rollover poses a major limiting factor in power scaling of VECSELs [5]. Most commonly, these lasers employ an integrated distributed Bragg reflector (DBR) as the back reflector of the external cavity. The relatively large thermal resistance of the DBR hinders efficient heat removal from the gain region at high excitation levels. As a remedy, diamond heat spreaders, soldered to the gain chip, have been used to obtain CW power exceeding 100 W, albeit with poor beam quality [6]. Most recently, DBR-free "membrane" VECSELs or membrane-external cavity surface-emitting lasers (MECSELs) have been introduced. They employ van der Waals bonding of the gain chip to heat spreader such as silicon carbide (SiC) or diamond [7][8][9][10][11]. This method shows promise on improving heat mitigation in VECSELs, provided that the intracavity heat spreader (preferably diamond) is of excellent optical quality. Nonetheless, the relatively small ring-shaped contact area between the heat spreader and the heat sink still limits the utility for high power operations [10]. To circumvent these limitations, we introduce novel active mirror structures with gain embedded in broadband subwavelength grating reflectors, which has low thermal resistance and allows near one dimensional heat flow.

Subwavelength grating structures have been studied for a wide range of integrated optoelectronic applications including narrow-band filters, lasers, couplers and broadband high reflectors [12]–[14][15]. In particular, using high-contrast refractive indices, broadband mirrors with reflectivity exceeding 99% have been demonstrated [16][17]. To explain the observed characteristics of these devices, various interpretations of Maxwell's equations along with detailed analysis based on the diffraction theory have been presented [8][9][20]. Overall, the narrow-band features exhibited in reflection or transmission (e.g. for filters) have been attributed to guided-mode resonances (GMR) involving leaky modes [19]. In simpler terms, the principal mechanism can be explained by a multi-beam interference process as follows: The incident wave diffracts into m=0, $\pm 1$ (and possibly but rarely higher-order) modes in the grating medium. These waves then experience multiple reflection/diffraction at the planar substrate and grating interfaces. In the latter, these modes couple or diffract into (or exchange energy among) each other. For a given wavelength and index contrast, there is always a grating geometry for which the transmitted waves in the substrate medium destructively interfere resulting in a small transmission which in turn implies a high reflectivity for the 0-th order in the incident medium (air). Such reflectivity bands show Fabry-Perot-like resonances (i.e. they are periodic with the roundtrip phase inside the grating medium) but are extremely narrow due to interference of multiple diffraction orders, and sensitive to the phase and grating geometry.

Our primary focus in this study concerns the broadband reflective structures [16][21], as they are potentially more tolerant to fabrication errors, and offer high heat-dissipating capabilities. Hereafter, we refer to all active mirror structures that use subwavelength gratings as "gain-embedded meta mirrors" or GEMMs.

The broadband high-reflectivity grating operates in a different regime compared with the narrow-band GMR grating. Like the narrow band grating, the broadband grating efficiently couples the incident (m=0) wave into non-zero orders (primarily m=±1) in the grating medium (see Fig. 1). These non-zero orders then experience total internal reflection (TIR) at the planar interface between high-index material and the low index substrate material, before coupling back into the 0-th order in the incident medium [16]. In comparison to narrow band GMR reflectors that show Fabry-Perot features, the high reflectivity devices are analogous to a Gires–Tournois interferometer where only the phase but not the amplitude of reflectance is modulated by the roundtrip phase. The bandwidth of these reflectors is then limited only by the wavelength range for which the diffraction efficiency into m=±1 modes in the grating medium remains near unity.

**Error! Reference source not found.** shows a simple 1D rectangular subwavelength grating structure with period $\Lambda$, height h, fill factor $f$, and index $n_H$. Optical plane waves with wavelength $\lambda_0$ are incident from above where the medium has refractive index $n_i$. For most practical applications (and in our analysis here) we take $n_i$=1 as in free space. The term "subwavelength" is used merely to imply that $\Lambda<\lambda_0/n_i$ thus allowing only m=0 diffraction order in the incident medium.

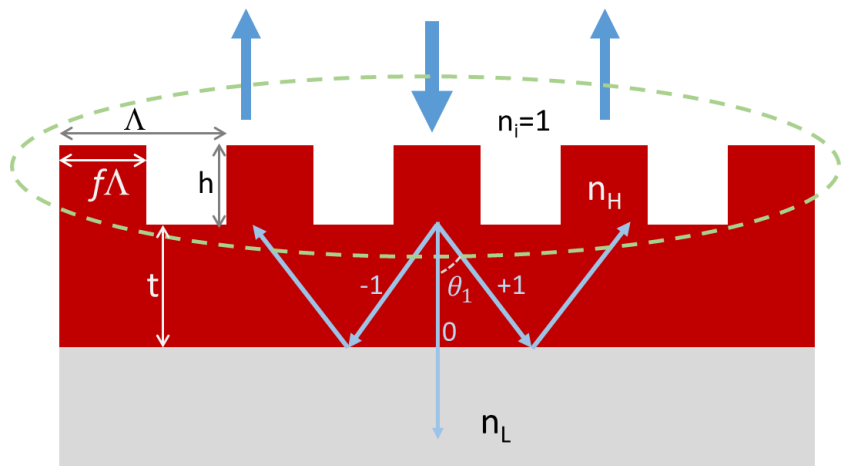

Figure 1 Schematic of subwavelength grating having a period $\Lambda$, height h, fill factor f and index of refraction $n_H$. The grating is bonded to a substrate of index $n_L< n_H$. The optical field under normal incidence is diffracted into m=0, ±1,.. orders while the subwavelength nature of the grating only allows m=0 reflection. The diffracted higher order modes will couple backward after TIR from the semiconductor/substrate interface.

By selecting the appropriate grating geometry and high index contrast ($n_H/n_i$), power can be coupled into m=±1 (or higher) diffraction orders in the grating medium with very high efficiency ($\eta_m$), thus leaving negligible power in the m=0 order ($\eta_0$~0). Taking the simplest and most practical case, we consider the grating period that only allows m=0, and ±1 orders, under the normal incidence condition. The diffraction angle is given by the grating equation: $\sin\theta_{\pm 1} = \pm\frac{\lambda_0}{\Lambda n_H} < 1$. The TIR condition at the planar interface with the substrate requires $|\sin\theta_{\pm 1}| > n_L/n_H$, which in turn leads to the condition

$$1 < \frac{\Lambda n_H}{\lambda_0} < n_H/n_L \quad (1)$$

that constrains the grating period for broadband reflectors [20][22].

The first task at hand is finding the parameters of the grating interface that minimize the transmitted first-order diffraction efficiency ($\eta_0$). Another approach is finding the parameters of the grating interface that maximize the transmitted first-order diffraction efficiency ($\eta_{\pm 1}$) while being least susceptible to small deviations due to fabrication errors and imperfections. Highlighted within the dashed-line ellipse in **Error! Reference source not found.**, these parameters are period $\Lambda$, fill factor $f$, grating depth $h$, and the grating index $n_H$. Note that the thickness $t$ does not play a critical role in the broadband regime. Such grating optimization, as has been reported by others, can be efficiently performed using the rigorous coupled-wave analysis (RCWA). Since this work is concerned with gain embedded structures that are most composed of III-V semiconductor compounds (e.g. multiple quantum wells, MQW), henceforth we will assume $n_H$=3.46, which applies to GaAs or compounds such as AlGaAs or GaInP. In the RCWA framework [23], one deals with the normalized parameters $\Lambda n_H/\lambda_0$, $f$ and $h/\Lambda$; therefore, we search this parameter space with particular interest in regions where $\eta_0$<0.01.

The left-hand side of this inequality enforces the subwavelength nature while the right-hand side ensures TIR at the substrate interface. Therefore, the choice of substrate materials and $n_H/n_L$ play an important role in the available range of parameters. Note also that for materials of interest in the present work, $n_H/n_L \lesssim 2$, only m=0, and ±1 orders are allowed in the grating medium. Once embedded with gain, we are ultimately concerned with efficient heat transport from the active area if high power operation is desired. This necessitates the use of substrates with a high coefficient of thermal conductivity $\kappa$ and good optical quality. In our analysis here, we specifically consider three such known candidates, namely diamond ($n_L \approx 2.4$, $\kappa \approx 2000\ W/m \cdot K$), SiC ($n_L \approx 2.5$, $\kappa \approx 340\ W/m \cdot K$), and sapphire ($n_L \approx 1.75$).

Figure 2 shows 2D example plots of $\eta_0$ versus $\Lambda n_H/\lambda_0$ and $h/\Lambda$ for a range of fill-factors $f$ that are calculated for TM polarization using the RCWA formalism (RSoft, Inc.) [23]. The region of interest ($\eta_0$<0.01) is marked (white area) for all graphs. Vertical lines in each graph signifies the lower bound (solid black line) and the upper bound (dashed blue lines) of the inequality in Eq. (1) for the three thermal substrates under consideration. For the index contrast ($n_H/n_i$=3.46) used, it is not totally unexpected that the lowest index substrate (here sapphire) has the highest margin for success while using SiC of the highest refractive index appears to be rather challenging. The general trend is that smaller fill factor favors lower index substrates. Fortunately, 0.55<$f$<0.70 offers a comfortable margin for using diamond as the thermal substrate which is of highest practical interest. Since either wet or dry etching is employed to generate the target grating grooves, which is most prone to fabrication variability, it is desirable to have a relatively broad range of etch depths (e.g. $\Delta h/h \geq 0.1$). We found that a similar favorable condition does not exist for TE modes, as has been the conclusion in earlier investigations of similar broad-band grating reflectors [20].

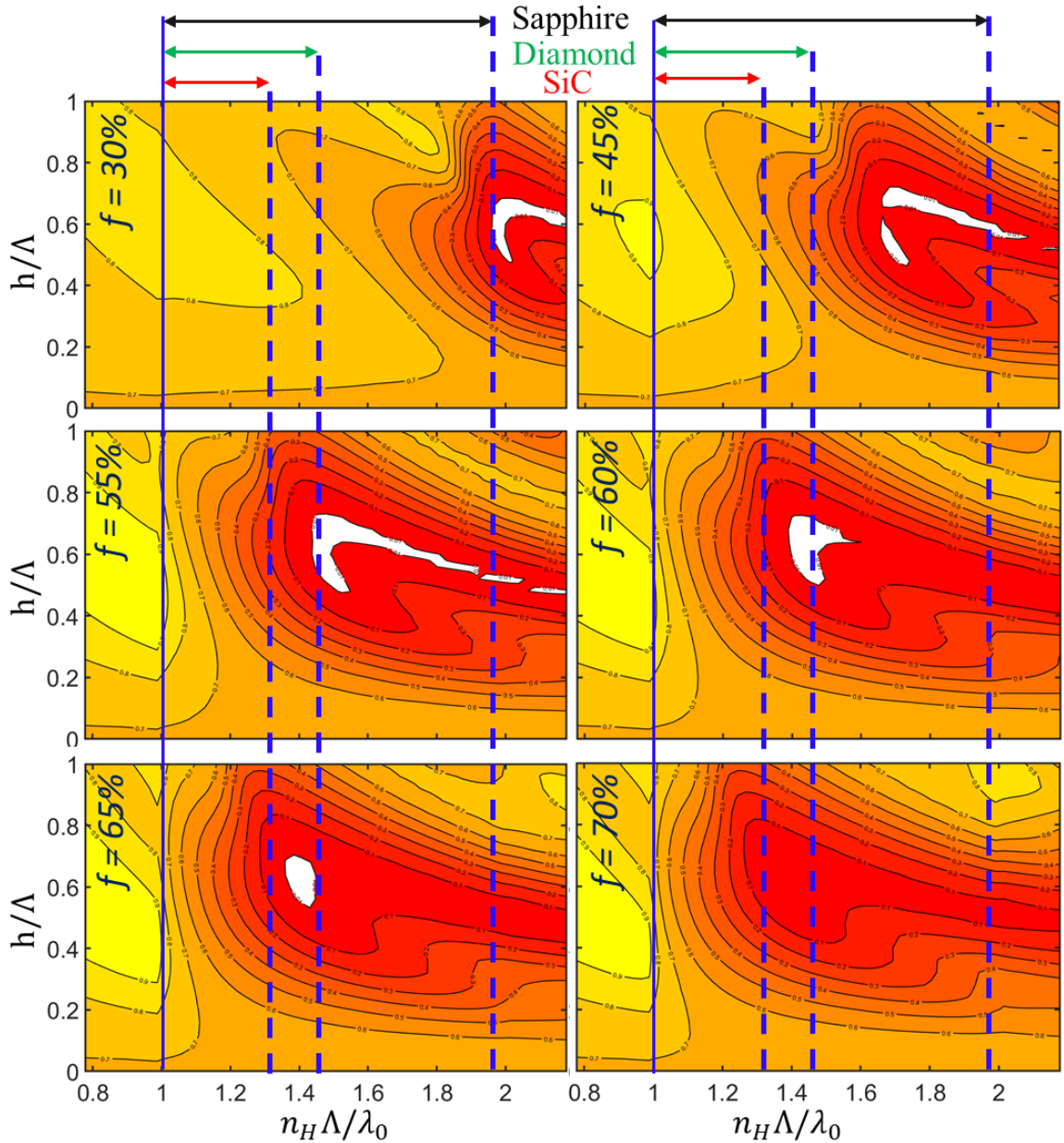

*Figure 2 Calcualted diffraction effiency $\eta_0$ of the incident optical field (TM polarization) into the m=0 transmitted mode versus normalized graing paramters $\Lambda n_H/\lambda_0$ and $h/\Lambda$ for various fill factors f. The white area corresponds to $\eta_0 < 1\%$. The vertical lines (solid and dashed blue) mark the limits of inequality in Eq. (1) for three thermal substrates of sapphire, diamond and SiC. For example, for a sapphire substrate, the possible values of $\Lambda n_H/\lambda_0$ range from 1 to 2. SiC and diamond have smaller design space. The black contour curves are at an increment of 10%.*

The calculated passive reflectivity maps of two such gratings versus the normalized design parameters are shown in Fig. 3 for two substrates of sapphire and diamond with fill factors *f = 55%* and *f = 63.5%* respctively.

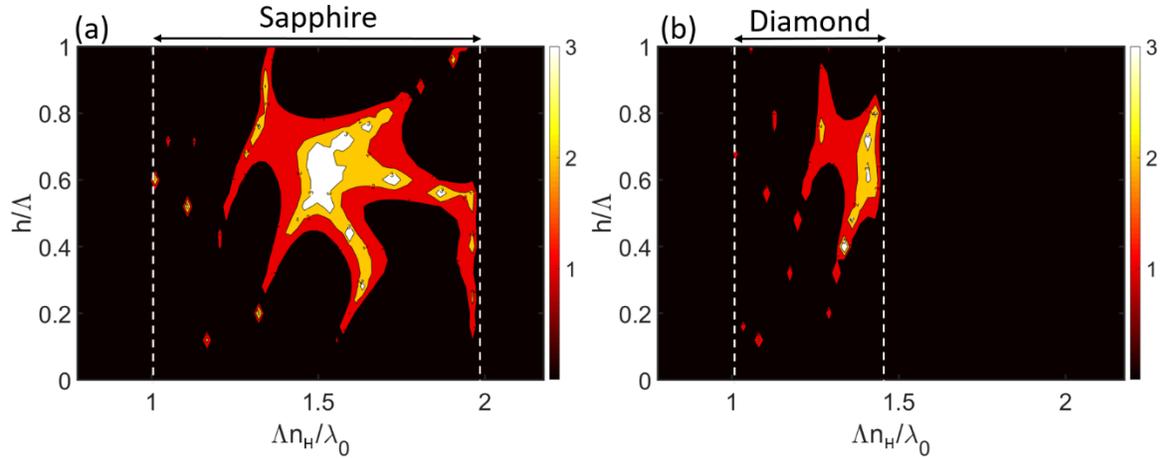

Figure 3. Calcualted passive reflectivity maps ( $\log_{10}\frac{1}{1-R}$ ) versus normalized grating pitch $n_H\Lambda/\lambda_0$ and height $h/\Lambda$ for two fill facotrs of (a) f = 55% and (b) f = 63.5% appropriate for sapphire and diamond thermal substrates respectively. The areas in white correspond to a reflectivity of 99.9%.

We will next incorporate optical gain into the grating medium. Since, as discussed earlier, broadband high reflection arises from TIR of m=±1 orders from the planar interface and negligible reflection from m=0 order at the grating interface, we do not encounter substantial subcavity resonances and local field enhancements. The field distribution is essentially determined by the interference of the +1 and -1 orders and their TIR counterparts, as shown in Fig. 4 for a GEMM reflector with Al$_{0.2}$Ga$_{0.8}$As ($n_H$=3.46) on diamond substrate having $n_H\Lambda/\lambda_0 = 1.35$ and $h/\Lambda$=0.6.

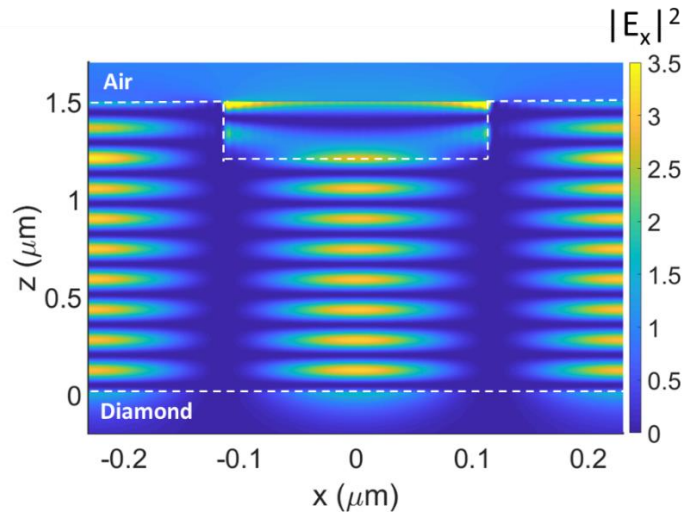

Fig. 4 The x-component of the electric-field distribution $|E_x|^2$ inside the grating medium calculated for f = 61%, $n_H\Lambda/\lambda_0 = 1.40$ and $h/\Lambda$ = 0.75. The white dashed lines represent the boundary of the grating. Only one unit-cell is present here and periodic boundary conditiosn are assumed along the x-direction.

Therefore, inserting a gain region of thickness d (as shown in the inset of Fig. 5) is expected to increase the reflectivity of the passive structure, $R_g$ to approximately $R_g e^{2\gamma_0 d/\cos(\theta_1)}$ where $\gamma_0$ is the small signal gain, and $\theta_1=\cos^{-1}(\lambda_0/n_H\Lambda)$ is the |m|=1 diffraction angle as illustrated in Fig. 1. If the gain medium consists of multiple quantum well (MQW) structures, then similar to standard VECSEL (or VCSEL) devices in standing wave cavities, gain is more efficiently extracted if QWs are placed at the antinodes, in the so-called resonant periodic gain (RPG) structure [24]. In the case of GEMM, QWs should be separated by $\lambda_0/2n_H\cos(\theta_1)$. However, since the GEMM structure involves TM modes, the gain enhancement factor of 2 obtained from using RPG is partially balanced by the $\cos(\theta_1)^2$ factor arising from the fact that QW selection rules favor gain for E-field polarizations parallel to the plane of wells (i.e. along x-axis). As the characteristics of such broad-band reflectors are not sensitive to the thickness $t$ of the grating medium, the number of quantum wells that can be imbedded in the grating is only limited by the pump power absorption considerations.

Using diamond as the thermal substrate ($n_s$=2.40), we calculate the GEMM reflectivity with inserted gain. For simplicity, we assume the gain medium is bulk GaAs having a thickness of 750 nm (GaAs) in a GaAs/AlGaAs double heterostructures with two $Al_{0.2}Ga_{0.8}As$ passivation layers each having a thickness of 750 nm. The grating is therefore taken to be etched in the AlGaAs barrier layer with an index $n_1$=3.46. For simplicity we ignore index variation between GaAs and AlGaAs. Following design parameters obtained from Fig. 2 for a diamond substrate, we choose $n_H\Lambda/\lambda_0$=1.40 and h/$\Lambda$=0.75, $f$=0.61 and take $\lambda_0$=890 nm corresponding to the peak of the gain of the GaAs. Assuming an optically injected electron-hole density $N_{eh}$=2.2 × $10^{18}$ cm$^{-3}$, we calculate the gain spectrum with a peak gain of 500 cm$^{-1}$ at 890nm using known semiconductor gain formalism [25]. This excitation level corresponds to an approximate steady-state absorbed power of 5.5 W at a pump wavelength of 808 nm assuming a pump beam radius of 200 μm. We also assumed a realistic case of elevated temperature of 370 K in calculating the optical gain in GaAs. The calculated reflection spectra of the GEMM structure under both passive and active (pumped) conditions are shown in Fig. 4. A reflectivity exceeding 115% is seen within a spectral range of 15 nm, limited only by the gain spectrum of the bulk GaAs active region. Embedding MQWs as the active layer results in a broader and flatter gain spectrum that can utilize the broad reflection spectrum of the passive GEMM shown in Fig. 3. This broadband gain can be exploited for continuous tunability [26] as well as modelocking of these lasers[27]–[29] .

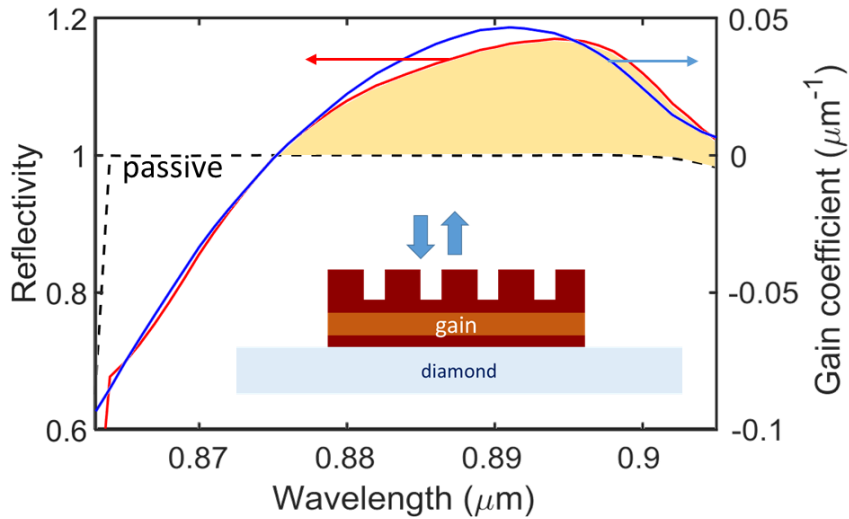

Figure 5 The spectra of reflectivity with (red solid) and without gain (black dashline), and gain coefficient (orange). The inset shows the GEMM structre bonded to a diamond substrate with an inserted active layer of thickness d.

We next evaluate the thermal performance of GEMM as an active mirror in a VECSEL device as depicted in Fig. 6a. We utilize the commercial software COMSOL based on the finite element method and compare the temperature rise between a GEMM-based and a traditional DBR-based VECSEL subject to a variable heat load deposited in the active region. Both the DBR- and GEMM-based VECSELs are mounted onto 0.5 mm thick diamonds, which are attached to a 1 mm thick copper mount with 20 µm thick indium foil. Substrates and samples are 8 mm in radius. The constant temperature (300 K) condition is applied to the bottom of either the copper mount or the diamond substrate. To reduce the computation time, we approximate the semiconductor region as two layers: cap (passive) and gain (active) where heat is deposited. DBR is also approximated as one layer [30]. The GEMM sample is directly bonded onto diamond, while the VECSEL sample is soldered with a 2 µm thick $AuIn_2$ solder layer (see the right panel of Fig. 6b). The values of thermal conductivity [30] used in the simulation are listed in Table 1:

Table 1. Material thermal conductivity values used in calculation

|  | Cap layer | Gain | DBR | Diamond | Copper | Indium | $AuIn_2$ solder |
|---|---|---|---|---|---|---|---|
| Thickness (µm) | 0.33, 0.36 | 1.5 | 4.3 | 500 | 1000 | 20 | 2 |
| κ (W/m·K) | 10 | 24 | 35 | 1800 | 400 | 84 | 30 |

With a pump spot radius of 5 mm, we simulate the maximum temperature rises in active regions under various heat loads in four heat-sinking schemes, as shown in the right panel of Fig. 6b. In the previous work on high-power VECSEL operation, with the DBR soldered to diamond (as in Fig. 6b), the thermal rollover was estimated to occur at a temperature rise of 100°C [31]. We have therefore used this value of ∆T as a point of comparison between the two structures in Fig. 6b. We see that the GEMM-based VECSEL can outperform traditional VECSEL by nearly a factor of 2. With a 5 mm spot size, the thermal model predicts that nearly 1.5 kW of heat load can be dissipated. As the slope efficiencies of these lasers can exceed 50% when pumped under low-quantum-defect conditions, we estimate that GEMM-based VECSELs can reach kilowatt output operation. However, this is primarily limited by the thermal resistance of the copper mount. That is, if we keep the bottom of the heat spreader (diamond) at a constant (heat sink) temperature, such as in water jet impingement scheme [32][33], the GEMM configuration outperforms the DBR based VECSEL by tolerating nearly 8 kW of heat load, corresponding to an enhancement of 3.2. Therefore, in addition to common advantages of a DBR-free VECSEL, the GEMM structure can potentially be exploited to power scale into the multi-kW power regimes without significant thermal degradation.

In conclusion, we have proposed and analyzed an active mirror concept based on a gain-embedded meta-mirror (GEMM) that uses a broad-band active subwavelength grating structure bonded to a thermal substrate. Using RCWA formalism, we identified the grating parameters that will work for high thermal conductivity substrates. In particular, we performed thermal analysis on a GaAs-based GEMM-on-diamond structure and showed that it can outperform traditional VECSELs (with DBR-on-diamond) by more than a factor of 3. An optically pumped semiconductor disk laser based on GEMM can therefore be used to push the power of these lasers into multi-kW continuous wave regimes. The GEMM structure additionally offers the desirable advantages of DBR-free (membrane) VECSELs including its potential for wide wavelength tunability and accessibility of spectral regions for which epitaxial DBRs are not readily available.

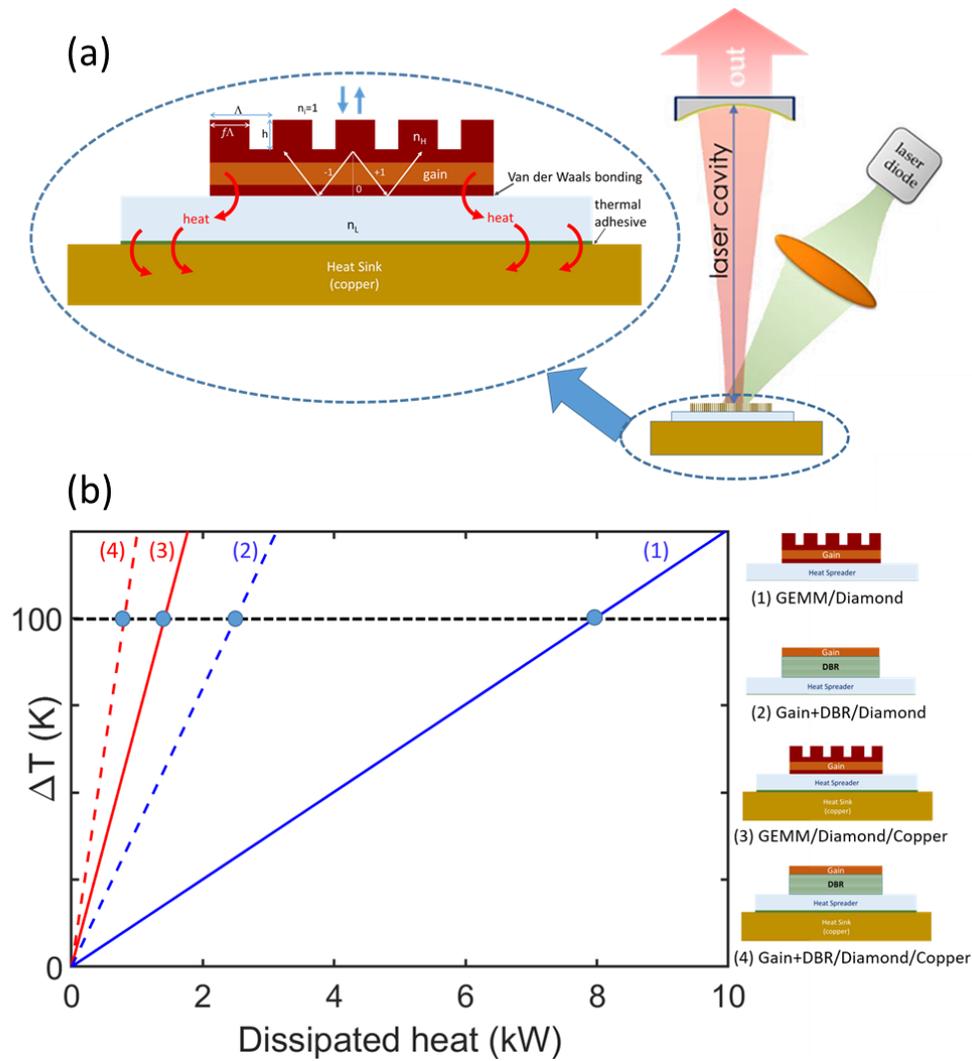

*Figure 6 (a) An example of GEMM used as an active mirror in a linear VECSEL cavity. (b) The maximum temperature rise in the gain region versus dissipated heat load for VECSELs based on (1) GEMM (solid lines) and (2) DBR (dashed lines) structures as depicted in the insets. The thermal substrate in both cases is taken to be .5 mm thick diamond. The heat load is taken to be inside the gain region within a circualr area assuming two pump spot sizes (radius) of 1 mm (red lines) and 5 mm (blue lines).*

**Acknowledgments:** This work was performed under funding from Air Force Office of Scientific Research (AFSOR) FA9550-16-0362 (MURI), and Air Force Research Laboratory (AFRL), Phase-I STTR. The authors also thank Alexander Albrecht for useful discussions. This work was performed, in part, at the Center for Integrated Nanotechnologies, an Office of Science User Facility operated for the U.S. Department of Energy (DOE) Office of Science by Los Alamos National Laboratory (Contract DE-AC52-06NA25396) and Sandia National Laboratories (Contract DE-NA-0003525).